\documentstyle[epsfig,floats,tighten,preprint,prd,aps]{revtex}

\begin{document}

\preprint{CITA-2002-09, hep-th/0204187}

\title{Prospects and Problems of Tachyon Matter Cosmology}

\author{Andrei Frolov}
\address{
  Canadian Institute for Theoretical Astrophysics,
  University of Toronto\\
  Toronto, ON, M5S 3H8, Canada
}
\author{Lev Kofman}
\address{
  Canadian Institute for Theoretical Astrophysics,
  University of Toronto\\
  Toronto, ON, M5S 3H8, Canada
}
\author{Alexei Starobinsky}
\address{
  Landau Institute for Theoretical Physics,
  Kosygina 2,\\
  Moscow, 117334, Russia
}

\date{April 22, 2002}

\maketitle

\begin{abstract}
  We consider the evolution of FRW cosmological models and linear
  perturbations of tachyon matter rolling towards a minimum of its
  potential. The tachyon coupled to gravity is described by an
  effective 4d field theory of string theory tachyon. In the model
  where a tachyon potential $V(T)$ has a quadratic minimum at finite
  value of the tachyon field $T_0$ and $V(T_0)=0$, the tachyon
  condensate oscillates around its minimum with a decreasing amplitude.
  It is shown that its effective equation of state is
  $p=-\varepsilon/3$. However, linear inhomogeneous tachyon
  fluctuations coupled to the oscillating background condensate are
  exponentially unstable due to the effect of parametric resonance. In
  another interesting model, where tachyon potential exponentially
  approaches zero at infinity of $T$, rolling tachyon condensate in an
  expanding universe behaves as pressureless fluid. Its linear
  fluctuations coupled with small metric perturbations evolve similar
  to these in the pressureless fluid. However, this linear stage
  changes to a strongly non-linear one very early, so that the usual
  quasi-linear stage observed at sufficiently large scales in the
  present Universe may not be realized in the absence of the usual
  particle-like cold dark matter.
\end{abstract}


\section{Introduction}

There are many faces of superstring/brane cosmology which come from
different corners of M/String theories. In particular, people search
for potential candidates to explain early universe inflation, present
day dark energy and dark matter in the universe. One of the string
theory constructions, tachyon on D-branes, has been recently proposed
for cosmological applications by Sen \cite{Sen2}. A relatively simple
formulation of the unstable D-brane tachyon dynamics in terms of
effective field theory stimulates one to investigate its role in
cosmology \cite{Gibb}.

\begin{figure}[t]
  \begin{center}
    \begin{tabular}{c@{\hspace{3cm}}c}
      \epsfig{file=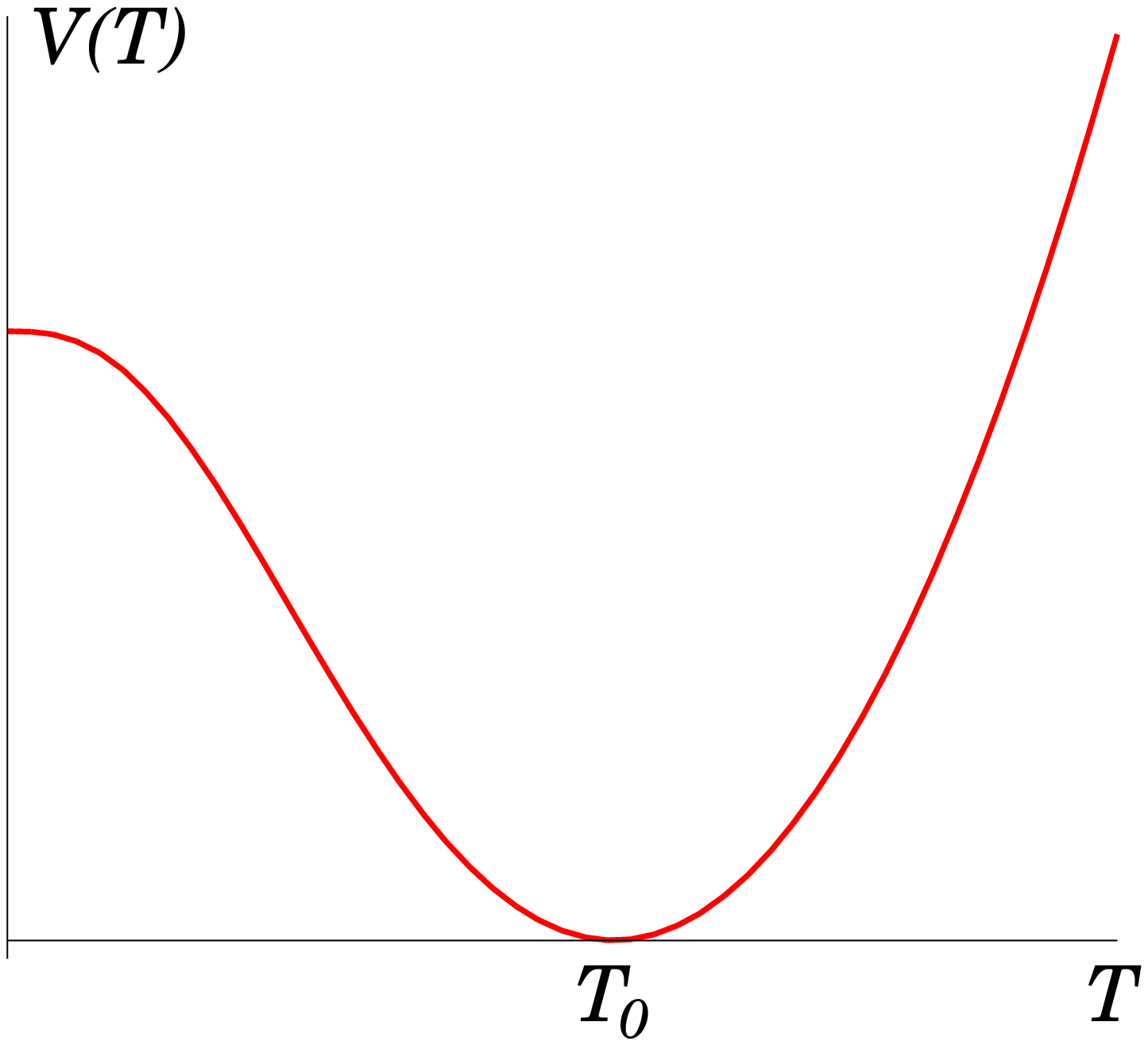, width=5cm} &
      \epsfig{file=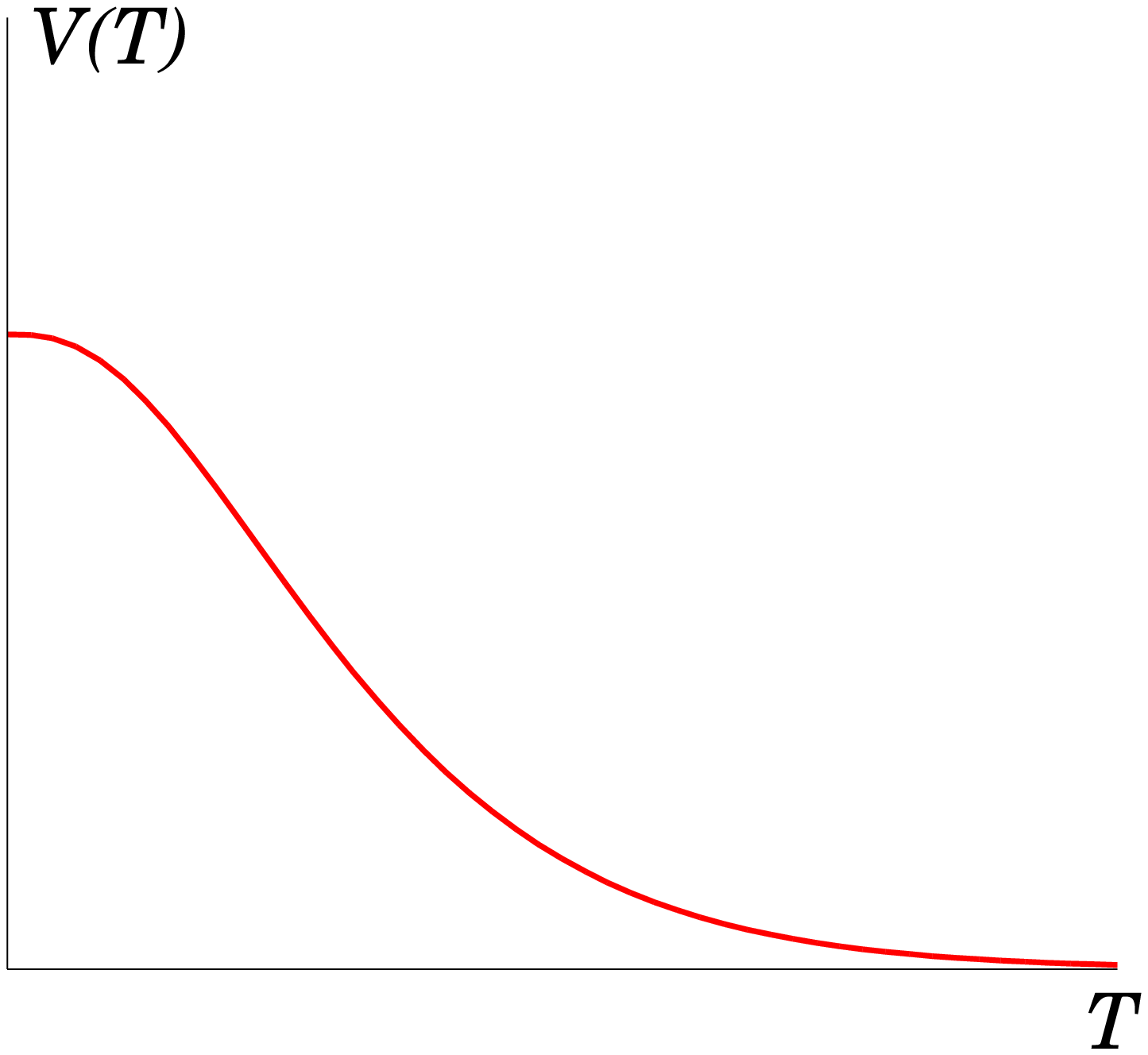, width=5cm} \\
      (a) & (b)\\
    \end{tabular}
  \end{center}
  \caption{
    Tachyon matter potentials with minima at finite (a) and infinite (b)
    values of the field. The potentials near minimum are taken to be:
    (a) $V(T) = \frac{1}{2} m^2 (T-T_0)^2$,
    (b) $V(T) = V_0 e^{-T/T_0}$.
  }
\label{fig:pot}
\end{figure}

The rolling tachyon in the string theory may be described in terms of
effective field theory for the tachyon condensate $T$, which in the
flat spacetime has a Lagrangian density
\begin{equation}\label{lag}
  {\cal L} = -V(T) \sqrt{1 + \partial_\mu T \partial ^\mu T} \ .
\end{equation}
The tachyon potential $V(T)$ has a positive maximum at $T=0$ and a
minimum at $T_0$, with $V(T_0)=0$. We consider two models, with finite
value of $T_0$ and with minimum at infinity, as illustrated in
Figure~\ref{fig:pot}. In both cases one encounters interesting
possibilities for cosmological applications.

In the case of finite $T_0$, we consider quadratic expansion around the
minimum of the potential $V(T) \approx {1 \over 2}m^2 (T-T_0)^2$. As we
will show, in this case the tachyon matter has negative pressure and
may be considered a candidate for quintessence.

In the case when $T_0 \to \infty$, we use exponential asymptotic of the
potential $V(T)=V_0 e^{-T/T_0}$ derived from the string theory
calculations \cite{exp,Sen1} (exact form of the potential from
Ref.~\cite{exp} is $V=(1+\frac{T}{T_0}) e^{-T/T_0}$; our qualitative
results for late time asymptotics of $T(t)$ do not depend on the
pre-exponential factor). Dimensional parameters of the potential are
related to the fundamental length scale, $T_0 \sim l_s$, and $V_0$ is
the brane tension. As it was demonstrated by Sen \cite{Sen1}, the
tachyon matter is pressureless for the potential with the ground state
at infinity. In this case tachyon matter may be considered a cold dark
matter candidate \cite{Sen2}.

It is noteworthy that the models of type (\ref{lag}) already were
studied in cosmology on the phenomenological ground. For certain
choices of potentials $V$ and non-minimal kinetic terms one can get
kinematically driven inflation, ``$k$-inflation'' \cite{k}. In
particular, a toy model with the potential $V(T) \sim 1/T^2$ with
ground state at infinity may give rise to the power law inflation of
the universe \cite{k,perk,alex}. However, it remains to be seen how
this potential can be motivated by the string theory of tachyon. The
model with $V \equiv \text{const}$ is reduced to the so-called
``Chaplygin gas'' where the matter equation of state is
$p=-\text{const}/\varepsilon$. Such matter was suggested as a candidate
for the present dark energy \cite{kam,fab}.

In this paper, we investigate cosmology with tachyon matter with the
string theory motivated potentials of Figure~\ref{fig:pot}. In
Section~\ref{sec:basics}, we write down the equations for the tachyon
matter coupled to gravity. We focus on self-consistent formulation of
the isotropic Friedmann-Robertson-Walker cosmology supported by the
tachyon matter. It is described by coupled equations for the
time-dependent background tachyon field $T(t)$ and the scale factor of
the universe $a(t)$.

One of the lessons of the scalar field theory in cosmology is the
possibility of the fast growth of inhomogeneous scalar field
fluctuations, as it was found in different situations. Instability of
scalar field fluctuations are typical for preheating after inflation
due to the parametric resonance \cite{KLS}, or tachyonic preheating
after hybrid inflation \cite{tach} (which so far has only remote
relation with string theory tachyon). Fluctuations can be unstable in
axion cosmology due to the parametric resonance \cite{GKS}. Therefore,
we address the problem of stability of linear fluctuations of the
rolling tachyon matter.

Consistent investigation of tachyon cosmology, in principle, should be
started with the tachyon rolling from the top of its potential, which
has negative curvature. In this case the setting of the problem is
similar to what we met in tachyonic preheating after hybrid inflation
\cite{tach}. In these cases we expect fast decay of the scalar field
into long-wavelength inhomogeneities. Here we assume that somehow
homogeneous tachyon rolls towards the minimum of its potential as
universe expands, and consider tachyon fluctuations at the latest
stages of its evolution.

In Section~\ref{sec:fluc}, we develop a formalism for treating small
fluctuations of the tachyon field $\delta T(t, {\vec x})$. It is
possible to extend theory of tachyon matter fluctuations by including
scalar metric fluctuations. This allows us to address the issue of
gravitational instability in tachyon cosmology. Applying this analysis
for specific tachyon potentials, we will see that instability of
tachyon fluctuations is essential for the whole story of tachyon
cosmology.

In Section~\ref{sec:negat}, we consider background cosmological
solutions for the tachyon potential with the ground state at finite
value $T_0$. We find that tachyon field is oscillating around the
minimum of its potential, while its equation of state (averaged over
oscillations) is $p=-{1 \over 3}\varepsilon$. Then in
Section~\ref{sec:osc1}, we check the stability of tachyon fluctuations
around this background solutions, and find that they are exponentially
unstable due to the parametric resonance.

In Section~\ref{sec:press}, we repeat the analysis for the model where
tachyon potential is exponential $V(T) \propto e^{-T/T_0}$ and its
ground state is at $T \to \infty$. In this case, background
cosmological solution corresponds to the pressureless tachyon with
energy density $\varepsilon \propto 1/a^3$. In Section~\ref{sec:press2},
we consider small inhomogeneous tachyon and metric fluctuations, and
find gravitational instability of fluctuations around the background
solution. Specifically, we find that for the pressureless rolling
tachyon linear approximation for fluctuations becomes insufficient very
early. We argue that the rolling tachyon dark matter scenario may have
difficulties in explaining gravitational clustering and the large scale
velocity flows.

\section{Cosmology with Rolling Tachyon Matter}\label{sec:basics}

Rolling tachyon is associated with unstable D-branes, and
self-consistent inclusion of gravity may require higher-dimensional
Einstein equations with branes. Still, in the low energy limit, one
expects that the brane gravity is reduced to the four dimensional
Einstein theory \cite{brane}.

In this section, we consider tachyon matter coupled with Einstein
gravity in four dimensions. Tachyon matter is described by the
phenomenological Lagrangian density (\ref{lag}), where derivatives are
covariantly generalized with respect to the metric $g_{\mu\nu}$,
$\partial_\mu \to \nabla_\mu$. We use the metric with signature
$(-, +, +, +)$. The model is given by the action
\begin{equation}\label{th}
S= \int d^4x\sqrt{-g}\, \left( { R \over 16 \pi
G} - V(T) \sqrt {1 + \nabla_\mu T \nabla^\mu T} \right) \ .
\end{equation}
The Einstein equations which follow from (\ref{th}) are
\begin{equation}\label{ein}
R_{\mu \nu}-\frac{1}{2} g_{\mu \nu }R= 8\pi G \left(
\frac{ V}{\sqrt {1 + \nabla_\alpha T \nabla^\alpha T }}\,\,
\nabla_\mu T \,\, \nabla_\nu T - g_{\mu \nu } V \,
\sqrt {1 + \nabla_\alpha T \nabla^\alpha T} \right) \ ,
\end{equation}
and the field equation for the tachyon is
\begin{equation}\label{field}
\nabla_\mu\nabla^\mu T -
\frac{\nabla_\mu \nabla_\nu T}
{1 + \nabla_\alpha T \nabla^\alpha T} \,\, \nabla^\mu T \,\, \nabla^\nu T
 - \frac{V_{,T}}{V}=0 \ .
\end{equation}
Let us apply these equations for the isotropic and homogeneous FRW
cosmological model
\begin{equation}\label{fr}
ds^2 = -dt ^2 + a^2(t) d {\vec x}^2 \ ,
\end{equation}
where $a(t)$ is the scale factor of the spatially flat ($K=0$)
universe. For this geometry, the energy-momentum tensor of tachyon
matter in the right-hand side of the equation (\ref{ein}) is reduced to
a diagonal form $T^{\mu}_{\nu}=\text{diag}(-\varepsilon, p, p, p)$,
where the energy density $\varepsilon$ is positive
\begin{equation}\label{density}
 \varepsilon = {V(T) \over \sqrt{1- \dot T ^2 }} \ ,
\end{equation}
and the pressure $p$ is negative or zero
\begin{equation}\label{pressure}
p = -V(T) \sqrt {1- \dot T ^2 } \ .
\end{equation}
Equation for the evolution of the scale factor follows from (\ref{ein})
\begin{equation}\label{Friedman}
{ \dot a^ 2 \over a^2 } = { 8\pi G
\over 3} { V(T) \over \sqrt { 1- \dot T^2 }} \ .
\end{equation}
Equation for the time-dependent rolling tachyon in an expanding
universe follows from (\ref{field})
\begin{equation}\label{time}
\frac{\ddot T }{1-\dot T^2}+ 3\, \frac{\dot a}{a}\, \dot T+ \frac{V_{,T}}{V} = 0\ .
\end{equation}
Note that the tachyon potential enters the field equation in a
combination $(\ln V)_{,T}$.

In the following sections we consider background solutions of equations
(\ref{Friedman}) and (\ref{time}) for two models of the tachyon
potentials $V(T)$ from Figure~\ref{fig:pot}.

\section{Fluctuations in Rolling Tachyon}\label{sec:fluc}

The issue of stability of small scalar fluctuations is often essential
in scalar field cosmology. In this section, we provide a formalism for
treating linear inhomogeneous fluctuations in the tachyon field. Let us
consider small inhomogeneous perturbation of the tachyon field $\delta
T(t, \vec{x})$ around time-dependent background solution $T(t)$ of
equation (\ref{time})
\begin{equation}\label{small}
T(t, \vec{x}) = T(t) + \delta T(t, \vec{x}) .
\end{equation}
As we will see, for one of our examples of tachyon potentials $V(T)$,
instability of tachyon fluctuations grows and becomes non-linear very
quickly. Therefore first we write down the equation for fluctuations
$\delta T(t,\vec{x})$ ignoring expansion of the universe and ignoring
coupling of tachyon fluctuations with the metric fluctuations.

Linearizing the field equation (\ref{field}) (without Hubble friction
term) with respect to small fluctuations $\delta T$ and performing
Fourier decomposition $\delta T( t,\vec{x}) = \int d^3k\, T_k(t)
e^{i{\vec{k} \vec{x}}}$ of the linear fluctuations, we obtain evolution
equation for the time-dependent Fourier amplitudes $ T_k(t)$
\begin{equation}\label{fluct1}
\frac{\ddot T_k }{1-\dot T^2}+ \frac{2\dot T \ddot T}{(1-\dot T^2)^2}\dot T_k
+\left[ k^2+(\log V)_{,TT} \right]T_k = 0\ .
\end{equation}

Next we consider tachyon fluctuations coupled with metric perturbations
in an expanding universe. Small scalar metric perturbations around
expanding isotropic cosmology can be written in terms of Newtonian
gravitational potential $\Phi(t,\vec{x})$ as
\begin{equation}\label{adiab}
ds^{2}=\left( 1+2\Phi \right) dt^{2}-\left( 1-2\Phi \right) a^{2}\left(
t\right) d{\vec x}^2 \ .
\end{equation}
Now we have to linearize the Einstein equations (\ref{ein})
and the field equation (\ref{field}) with respect to small
fluctuations $\delta T$ and $\Phi$.

Fortunately, the formalism for cosmological scalar fluctuations for the
class of models which includes the theory (\ref{th}) was developed in
Ref.~\cite{k} (in connection with ``$k$-inflation''). This is exactly
what we need to pursue the investigation of small cosmological
fluctuations with tachyon matter. Using results of \cite{k}, from
(\ref{ein}) and (\ref{field}) we obtain two coupled equations for the
time-dependent Fourier amplitudes $ T_k(t)$ and $\Phi_k(t)$,
\begin{equation}\label{e1}
\left( \frac{T_k }{\dot T}\right) ^{.}=\left( 1-\frac{1}{4\pi G} \,\, \frac{k^2}{a^2}\,\,
\frac{ (1-\dot T^2)^{3/2}}{V \dot T^2 }\right) \Phi_k \ ,
\end{equation}
and
\begin{equation}\label{e2}
\frac{\left( a\Phi_k \right)^{.}}{a}=4\pi G \,\, \frac{V \dot T^2}{ (1-\dot T^2)^{1/2}}
 \,\, \frac{T_k }{\dot T} \ .
\end{equation}
Introducing Mukhanov's variable $v_k$, which is related to the
potential $\Phi_k$ as
\begin{equation}\label{mukh}
\frac{v_k}{z} =\frac{5\varepsilon +3p}{3\left( \varepsilon +p\right) }\Phi_k
 +\frac{2}{3} \frac{\varepsilon }{\varepsilon +p}\frac{\dot{\Phi_k}}{H} \ ,
\end{equation}
where energy density $\varepsilon$ and pressure $p$ are given by
equations (\ref{density}) and (\ref{pressure}), $H=\frac{\dot
a}{a}$, and
\begin{equation}\label{z}
z=\frac{\sqrt{3} a \dot T}{(1-\dot T^2)^{1/2}} \ ,
\end{equation}
equations (\ref{e1}) and (\ref{e2}) can be reduced to a single second
order equation for $v_k$
\begin{equation}\label{single}
v_{k}^{\prime \prime }+\left( (1-\dot T^2) k^{2}-\frac{z^{\prime \prime }}{z}%
\right) v_{k}=0 \ ,
\end{equation}
where prime (${'}$) stands for derivative with respect to the conformal
time $d\eta=dt/a(t)$. We will use this equation for analysis of coupled
tachyon and metric fluctuations in an expanding universe for the
tachyon potential $V(T) \propto e^{-T/T_0}$.

\section{Negative-Pressure Tachyon Matter}\label{sec:negat}

In this section, we consider the model with the potential $V(T)$ with
its ground state at the finite value $T_0$, as it is sketched in the
left panel of Figure~\ref{fig:pot}. Let us assume that tachyon is
rolling towards the minimum of the potential $T_0$. We will approximate
the shape of the tachyon potential around the minimum by a quadratic
form $V(T) \approx {1 \over 2}m^2 (T-T_0)^2$. Despite quadratic form of
the potential, tachyon motion around $T_0$ is not harmonic, since $\ln
V$ but not $V$ is involved in the tachyon equation of motion.
For the same reason parameter $m$ drops out of the field equation
(\ref{time}). It is convenient to use tachyon field in units of $T_0$
and time $t$ also in units of $T_0$. Parameter $m$, however, is
involved in the energy density of tachyon $\varepsilon \propto
\frac{m^2} {t^2}$. The choice of $m \sim l_s^{-1} \sim M_p$ may bring
the value of $\varepsilon$ to the required density of dark energy.

\begin{figure}[t]
  \begin{center}
    \begin{tabular}{c@{\hspace{3cm}}c}
      \epsfig{file=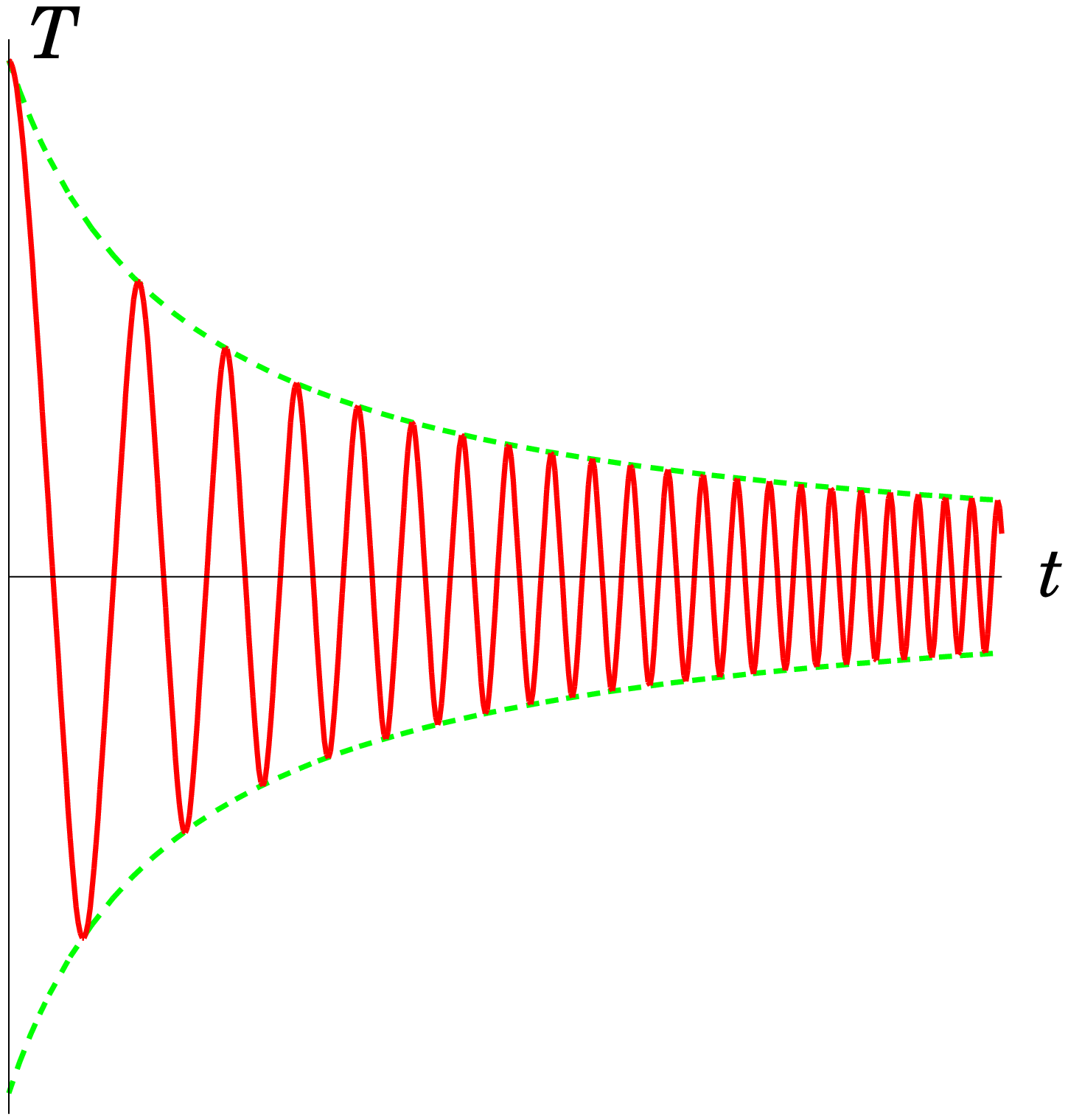, width=6cm} &
      \epsfig{file=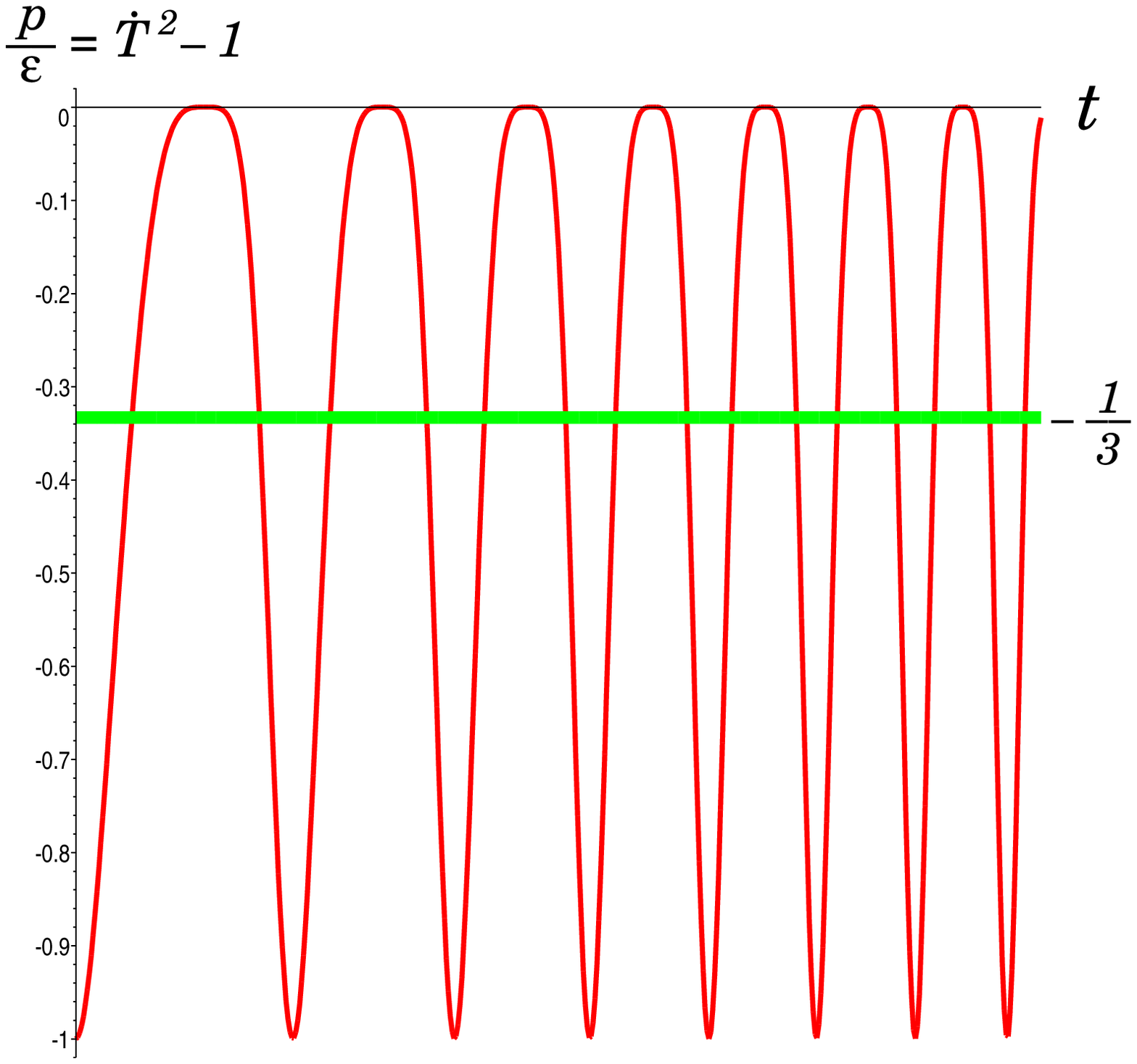, width=6cm} \\
      (a) & (b)\\
    \end{tabular}
  \end{center}
  \caption{
    (a) Background tachyon oscillations in the model with $V(T) =
    \frac{1}{2} m^2 (T-T_0)^2$. (b) Background oscillations of tachyon
    equation of state. Constant horizontal line $\frac{p}{\varepsilon}
    = -\frac{1}{3}$ is the time-averaged equation of state.
  }
  \label{fig:eos}
\end{figure}

Numerical solution of the equations (\ref{time}) and (\ref{Friedman})
reveals that tachyon very quickly, within time interval of several
$T_0$, begins to oscillate around the minimum of the potential, as
shown in the left panel of Figure~\ref{fig:eos}. The amplitude of the
oscillations is decreasing with time due to the Hubble friction term in
the equation (\ref{time}). The envelope curve (dashed line) in the left
panel of Figure~(\ref{fig:eos}) shows the amplitude decreasing as
$1/t$. As we will see below, this time-dependence of the amplitude
exactly corresponds to the (time-averaged) equation of state
$\varepsilon/p$ which will be found for the tachyon matter in this
model. Also, note that the period of oscillations is decreasing with
time. Tachyon oscillations in this model are not only non-harmonic, but
also non-periodic.

The instant value of the ratio of energy density (\ref{density}) and
pressure (\ref{pressure}), ${p \over \varepsilon}=\dot T^2-1$, is
oscillating with time, as shown in the right panel of
Figure~\ref{fig:eos}. Although the amplitude of oscillations $T$ is
decreasing with time, the amplitude of $\dot T$ is not changing with
time as it is clear from the Figure~\ref{fig:eos}.

The period of oscillations is very small ($\sim T_0$), so that only the
average equation of state is important for cosmological evolution. To
find it, we average $\dot T^2-1$ over several consecutive oscillations.
The average value of ${p \over \varepsilon}$, shown as the horizontal
line at the right panel of Figure~\ref{fig:eos}, is independent of time
and equal to
\begin{equation}\label{eos}
\langle {p \over \varepsilon}\rangle =-\frac{1}{3} \ .
\end{equation}
For this type of the equation of state, average value of the energy
density dilutes as $\langle {\varepsilon} \rangle \propto a^{-2}$ with the
expansion of the universe. The amplitude of the tachyon oscillations is
then decreasing as $1/t$, which is compatible with numerical results.
From (\ref{Friedman}) we find that the scale factor on average is
$a(t) \propto t$. Note that equation of state similar to (\ref{eos}) takes
place for a network of cosmic strings.

Equation of state (\ref{eos}) for the quadratic tachyon potential can
be easily derived analytically. Indeed, assuming that tachyon is
oscillating much faster than the universe expands, we can treat energy
density $\varepsilon$ as adiabatic invariant, and write $\dot
T^2=1-{{V^2(T)} \over \varepsilon^2}$, where $\varepsilon$ is constant
over several consecutive oscillations. Then the average value of $\dot
T^2$ for quadratic potential is
\begin{equation}\label{aver}
\langle \dot T^2 \rangle = { {\int \dot T^2 \, dt } \over {\int dt}}=
{ {\oint (1-V^2(T)/\varepsilon^2)^{1/2}\, dT}
\over {\oint (1-V^2(T)/\varepsilon^2)^{-1/2}\, dT} }
=\frac{2}{3} \ ,
\end{equation}
and so $\langle {p \over \varepsilon}\rangle= \langle \dot T^2
\rangle-1=-\frac{1}{3}$. If shape of the potential around the minimum
is not quadratic, but a power-law $V \propto (T-T_0)^{n}$, the average
equation of state is $\langle {p \over \varepsilon}\rangle =-\frac{1}{n+1}$.

Although tachyon matter in the model has negative pressure, apparently
it is short of explaining present acceleration of the universe.
Combination of cosmological observations of CMB fluctuations, large
scale structure clustering and high redshift supernovae constrains the
equation of state to be lower than $\langle {p \over \varepsilon}
\rangle < -0.6$ \cite{obs}.

As we will see in the next section, background tachyon dynamics in this
model is unstable under small spatial fluctuations, and homogeneous
tachyon oscillations will decay. It will be interesting to find what
will be the final configuration of tachyon matter in this model and
what may be its potential application to cosmology.

\section{Fluctuations in Tachyon Matter with Negative Pressure}\label{sec:osc1}

\begin{figure}[t]
  \centerline{\epsfig{file=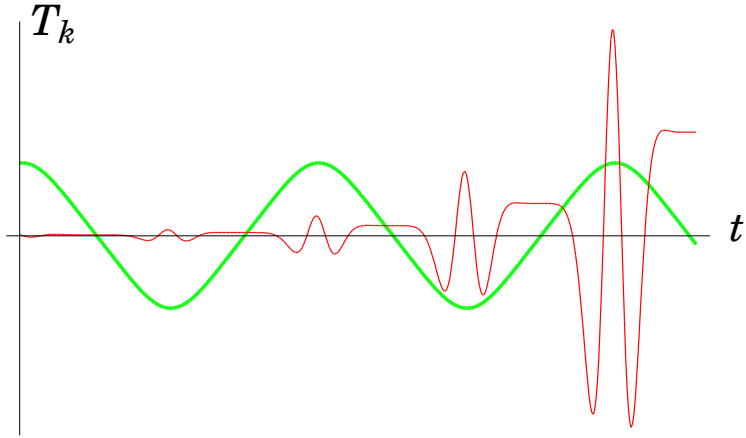, width=10cm}}
  \caption{
    Instability of fluctuations $T_k(t)$ in the model with quadratic
    potential (scales are linear).
  }
  \label{fig:inst}
\end{figure}

It is expected that in realistic cosmological scenario tachyon field
has small, quantum or classical, inhomogeneous fluctuations. In this
section, we check the stability of tachyon fluctuations around the
background solution discussed in the previous section. For the moment,
let us ignore the expansion of the universe. Then we only have to solve
equation (\ref{fluct1}) to find behaviour of fluctuations. Although
formally some of the coefficients in equation (\ref{fluct1}) are
singular when the background field $T(t)$ crosses zero in the case of
quadratic potential, it is possible to switch to regular variables and
overcome this technical inconvenience. Numerical solution of the
fluctuation equation (for example for $k=10$ in units of $T_0$) is
shown in Figure~\ref{fig:inst}.

General theory of linear equations with periodic coefficients predicts
the presence of stability and instability bands of momenta $k$. For
unstable modes, the amplitude is increasing exponentially as $T_k(t)
\sim e^{\mu_k t}$. For the value of $k$ in Figure~\ref{fig:inst} the
amplitude of fluctuations is increasing with time exponentially fast,
by an order of magnitude in one background oscillation, say $T_k(t)$
increases by a factor of $10^{10}$ in ten oscillations! The physical
reason is amplification due to the parametric resonance. This can be
clearly seen if one rewrites equation (\ref{fluct1}) in the form of the
oscillator-like equation, where the effective frequency is oscillating
with time. This effect can be described by the theory of broad
parametric resonance \cite{KLS97}.

Since period of oscillation ($\sim T_0$) is tiny compared to the
cosmological time, and fluctuations become significant within several
background oscillations, one can ignore expansion of the universe in
this analysis. Thus we conclude that in the model with quadratic
potential small tachyon fluctuations are exponentially unstable and
background tachyon condensate decays into inhomogeneous configuration.

Decay of background tachyon condensate into inhomogeneous fluctuations
does not necessarily mean that the universe becomes inhomogeneous. The
tachyon fluid remains homogeneous as a whole, but not as the
coherent condensate. Therefore it remains to be seen, based on the
fully nonlinear analysis, what will be equation of state of the
non-condensate tachyon fluid.

\section{Pressureless Tachyon Matter}\label{sec:press}

In this section, we consider the model with the potential $V(T)=V_0
e^{-T/T_0}$, with its ground state at infinity, as sketched in the
right panel of Figure~\ref{fig:pot}. Again, we can use dimensionless
tachyon $T$ in units of $T_0$ and dimensionless time $t$ in the same
units $T_0$. Background cosmological solutions of the equations
(\ref{time}) and (\ref{Friedman}) for this case very quickly (several
$T_0$) enters the regime where tachyon is rolling very fast, and $\dot
T$ approaches unity exponentially quickly
\begin{equation}\label{f1}
T(t)=t+ \frac{1}{4}a^6 e^{-{2t}} \ , \,\,\,\, \dot T^2=1-a^6 \, e^{-2t} \ ,
\end{equation}
where $a(t)$ is the scale factor. The formula (\ref{f1}) applies, in
particular, for radiation dominated stage where $a(t)=\sqrt{t}$ and
tachyon does not contribute to gravity, and for $a(t)=t^{2/3}$ where
tachyon gravitationally dominates. Without expansion of the universe,
solution (\ref{f1}) corresponds to that of Sen \cite{Sen2}.

In the above regime, tachyon potential is decreasing exponentially fast
$V(T) \sim e^{-t/T_0} $. Remarkably, as it was discovered by Sen
\cite{Sen2,Sen1}, the energy density of tachyon matter is non-zero,
while its pressure vanishes. In an expanding universe from
(\ref{density}), (\ref{pressure}), (\ref{f1}) we have
\begin{equation}\label{f2}
\varepsilon= \frac{V_0 \, e^{-t/T_0}}{a^3 \, e^{-t/T_0}} \to \frac{V_0}{a^3} \ ,
\end{equation}
while pressure vanishes
\begin{equation}\label{f3}
p=-V_0 \, e^{-t/T_0} \, a^3 e^{-t/T_0} \sim -a^3 e^{-2t/T_0} \to 0 \ .
\end{equation}
In dimensional units one can estimate $\varepsilon \sim M_p^2/t^2$. In
other words, rolling tachyon matter in an expanding universe is
pressureless, its energy density dilutes as $\varepsilon \propto
a^{-3}$, and its density may be tuned to the level of dark matter
density of the universe. This makes tachyon matter with exponential
potential a cosmological dark matter candidate.

\section{Cosmological Fluctuations for Pressureless Tachyon}\label{sec:press2}

The crucial property of the cosmological dark matter without pressure
is the growth of cosmological fluctuations, which form a developed
large scale structure. The large scale structure of the universe is
ranging from non-linear clustered halos of galaxies and clusters of
galaxies, quasi-linear structures at scales of supercluster and voids,
and linear fluctuations at very large scales. It is essential that at
quasi-linear and non-linear stages dark matter is displaced from the
homogeneous distribution due to the flows generated by the fluctuations
of gravitational potential, and gravitationally bound halos have high
velocity dispersions.

In this section, we investigate the clustering properties of the
pressureless tachyon matter. We assume rolling tachyon matter
domination and expansion rate of the universe is $a(t) \propto
t^{2/3}$.

We begin with linear analysis of cosmological fluctuations, using
formalism of Section~\ref{sec:fluc}. For a moment, consider the case
without expansion of the universe and without coupling to gravitational
perturbations. From equation (\ref{fluct1}) for exponential potential
it follows that the fluctuations are not growing, $T_k=\text{const}$.

Now let us consider tachyon fluctuations including expansion of the
universe and coupling to gravitational potential $\Phi$. Substituting
the background solution (\ref{f1}) for the pressureless tachyon
matter into equations (\ref{single}), one can see that the coefficient
in the front of $k^2$ (which plays the role of the sound speed for the
tachyon matter) vanishes exponentially fast. This means that the growth
of linear tachyon fluctuations is scale free, similar to that of the
standard cold dark matter scenario. Then the solution of (\ref{single})
is $v_k=z$, and the left-hand side of equation (\ref{mukh}) is
constant. From this we immediately get the time evolution of
fluctuations $\Phi_k$ and $T_k$
\begin{equation}\label{const}
\Phi_k(t)=\text{const} \ , \hspace{1em} T_k(t) = \Phi_k \cdot t \ .
\end{equation}
Linear metric fluctuations are constant, similar to that in the cold
dark matter scenario. However, the fluctuations in the tachyon field
are growing, $\dot T_k=1$, in contrast to the simplified analysis
above, where we neglected coupling to the metric fluctuations and
expansion of the universe. The growth of tachyon fluctuations $T_k
\propto t$ cannot be obtained without these ingredients. Thus, tachyon
fluctuations are unstable due to the effects of gravitational
instability in an expanding universe.

However, the linear approximation for rolling tachyon/gravity system
works only during very short time interval (of order of tens of $T_0$).
Indeed, let us inspect the energy density of tachyon matter in the
model with exponential potential, not assuming it is homogeneous
\begin{equation}\label{density1}
 \varepsilon = {{V_0 e^{-T/T_0}} \over \sqrt{1-\dot T^2+( \nabla_{\vec x}T)^2}} \ .
\end{equation}
For fluctuations $\delta T$ we have $\delta T \simeq \Phi(\vec x)\, t$,
where $\Phi(\vec x)$ describes initial spatial profile of the
fluctuations. The full tachyon field including fluctuations is $T(t,
\vec x)=(t+\frac{1}{4}a^6 e^{-2t}) + \Phi(\vec x) \,t$. The numerator
of expression (\ref{density1}) vanishes as $e^{-t}$, while the
denominator evolves as $a^6 e^{-2t}-2\Phi+(\nabla_{\vec x} \Phi)^2
t^2$. The linear approximation works during very short time interval
while $\left(\nabla_{\vec x} \Phi \right)^2 \lesssim e^{-2t}$. When
this inequality breaks, linear analysis becomes insufficient. For
cosmological fluctuations $\Phi \sim 10^{-5}$, linear theory is valid
during time interval of order of $10\, T_0$. Recall that for the
standard particle dark matter scenario linear stage lasts during
significant fraction of cosmological expansion.

One can try to use non-linear ansatz for the inhomogeneities of $T(t,
\vec x)$. Non-vanishing energy density (\ref{density1}) for rolling
tachyon with numerator $V\sim e^{-t}$ is achievable if denominator is
also vanishing as $e^{-t}$. Denominator $\sqrt{1- \dot T^2+
(\nabla_{\vec x} T)^2 }$ is not going to vanish unless we admit that
$(\nabla_{\vec x} T)^2 $
decreases faster than
$\sim e^{-2t}$. One of the working
ansatz for inhomogeneous tachyon field is
\begin{equation}\label{anz}
\ T\left(t, {\vec x } \right)= t+f({\vec x, t})e^{-2t} \ ,
\end{equation}
where $f({\vec x, t})$ weakly depends on $t$. This solution gives
energy density (\ref{density1}) which is only spatially dependent.
Solution for inhomogeneous tachyon was considered by Sen \cite{Sen3}.

Now we will argue that fast rolling tachyon condensate with vanishing
$p$ and nonzero $\varepsilon$ lacks nonvanishing gradients
$\nabla_{\vec x} T$ which correspond to cosmic velocity, or
displacement of tachyon matter. Indeed, energy-momentum tensor of
tachyon condensate may be written in the hydrodynamic form
\begin{equation}\label{hydro}
T_{\mu \nu }=(\varepsilon +p)u_{\mu }u_{\nu }+pg_{\mu \nu } \ ,
\end{equation}
where corresponding fluid velocity is
\begin{equation}\label{veloc}
u_{\mu }=\frac{\nabla_{\mu} T}{ (- \nabla_{\alpha}T \nabla^{\alpha}T)^{1/2}} \ ,
\end{equation}
energy density is given by (\ref{density1}) and pressure is $p= -V(T)
\sqrt {1+\nabla_{\alpha}T \nabla^{\alpha}T}$.

Substituting here the solution (\ref{anz}), we find $u_{\vec x} \sim
e^{-2t}$. Thus, rolling tachyon condensate does not behave like a media
of non-interacting particles. This is in contrast with the standard
pressureless media of cold dark matter particles where large scale
velocities and displacements of particles from their initial positions
are growing with time.

\section{Summary}\label{sec:summ}

We considered cosmological solutions of rolling tachyon condensate $T$
for two models of tachyon potential $V(T)$. There are different levels
at which one can theoreticize about $T(t)$ in expanding universe.
Systematic approach suggests for us to begin with theory of tachyon
field rolling down from the top of its potential. The curvature of the
potential at the origin is negative, and we expect tachyonic
instability of long wavelength fluctuations, similar to spinodal
instability in usual field theory \cite{tach}. Thus there is an issue
of initial conditions for rolling tachyon cosmology.

Suppose (by choice of initial conditions where $T$ is displaced from
the origin) tachyon evolves towards its ground state as homogeneous
condensate. We considered the model with minimum at finite $T_0$ and
with quadratic approximation of $V(T)$ around the minimum. Then the
background tachyon oscillates around the minimum with frequency of
order of $1/T_0$, and its (averaged over several oscillations) equation
of state is ${p \over \varepsilon}=-{1 \over 3}$. However, we found
from perturbation theory that tachyon fluctuations are exponentially
unstable due to
the tachyon self-interaction with background oscillations. It means
that in this model homogeneous tachyon condensate decays. To answer the
question what will be the resulting tachyon configuration, one has to
go to the next level beyond the perturbation theory and to consider
fully nonlinear problem of evolution of non-condensate
tachyon fluid.

We also consider homogeneous tachyon rolling towards its ground state
in the model with $V(T) \propto e^{-T/T_0}$, including expansion of the
universe. In this model, the background tachyon condensate has
vanishing pressure and finite energy density diluting as $\varepsilon
\propto 1/a^3$. Considering linear perturbations of tachyon field coupled
with small metric perturbations, we found gravitational instability of
tachyon field, $\delta T \propto t$. However, linear theory very soon (tens
of $T_0$) becomes irrelevant. We argue that non-vanishing energy
density of tachyon condensate is incompatible with large scale tachyon
flows. Again, to make more certain conclusion about tachyon cosmology
in this model, one has to use non-linear analysis to follow further
evolution of inhomogeneities.

\section*{Acknowledgements}

We are grateful to D.~Bond and A.~Linde for useful comments. We thank
NATO Linkage Grant 975389 for support. A.F. was supported by NSERC,
L.K. was supported by NSERC and CIAR. The work of A.S. in Russia was
partially supported by RFBR, grants No 02-02-16817 and 00-15-96699,
and by the RAS Research Program ``Astronomy''. A.S. thanks CITA for
hospitality during his visit.

\global\preprintstyfalse

\end{document}